\documentclass[reprint,amsmath,amssymb,aps,pra,superscriptaddress,floatfix]{revtex4-2}
\usepackage[utf8]{inputenc}
\usepackage[english]{babel}
\usepackage{hyperref}
\usepackage{dcolumn}
\usepackage{bm}
\usepackage{hyperref}
\usepackage{physics}
\usepackage[dvipsnames]{xcolor}
\usepackage{graphicx}
\usepackage[capitalise]{cleveref}
\usepackage{gensymb}
\usepackage{textcomp}
\usepackage{tabularx, cellspace}
\usepackage{bold-extra}
\usepackage{mhchem}
\usepackage{tabularx}
\usepackage[
    protrusion=true,
    activate={true,nocompatibility},
    final,
    tracking=true,
    kerning=true,
    spacing=true,
    factor=1000]{microtype}

\usepackage[scr=esstix,cal=boondoxo]{mathalfa} 
\usepackage[normalem]{ulem}

\definecolor{KartikRed}{RGB}{209, 0, 11}
\definecolor{gregGreen}{RGB}{205, 30, 214}

\newcommand{\ks}[1]{\textcolor{KartikRed}{#1}}

\newcommand{\gregsout}{\bgroup\markoverwith{\textcolor{gregGreen}{\rule[0.4ex]{2pt}{2pt}}}\ULon}

\hypersetup{
 pdftitle={},    
 pdfauthor={Gregory Moille},     
 pdfcreator={Gregory Moille},   
 pdfproducer={LaTeX}, 
 pdfkeywords={Version 1.0},
 colorlinks=true,       
 linkcolor=blue,          
 citecolor=blue,        
 filecolor=blue,      
 urlcolor=blue}

\usepackage{titlesec}
\titleformat*{\section}{\bfseries\Large}
\titleformat*{\subsection}{\bfseries}
\usepackage{caption}

\begin{document}
\author{Gr\'egory Moille}
\email{gmoille@umd.edu}
\affiliation{Joint Quantum Institute, NIST/University of Maryland, College Park, USA}
\affiliation{Microsystems and Nanotechnology Division, National Institute of Standards and Technology, Gaithersburg, USA}
\author{Christy Li}
\affiliation{Joint Quantum Institute, NIST/University of Maryland, College Park, USA}
\author{Jordan Stone}
\affiliation{Joint Quantum Institute, NIST/University of Maryland, College Park, USA}
\affiliation{Microsystems and Nanotechnology Division, National Institute of Standards and Technology, Gaithersburg, USA}
\author{Michal Chojnacky}
\affiliation{Joint Quantum Institute, NIST/University of Maryland, College Park, USA}
\author{Pradyoth Shandilya}
\affiliation{University of Maryland at Baltimore County, Baltimore, MD, USA}
\author{Yanne K. Chembo}
\affiliation{Institute for Research in Electronics and Applied Physics, University of Maryland,College Park, MD, USA}
\author{Avik Dutt}
\affiliation{Department of Mechanical Engineering and Institute for Physical Science and Technology, University of Maryland, College Park, MD 20742, USA}
\author{Curtis Menyuk}
\affiliation{University of Maryland at Baltimore County, Baltimore, MD, USA}
\author{Kartik Srinivasan}
\affiliation{Joint Quantum Institute, NIST/University of Maryland, College Park, USA}
\affiliation{Microsystems and Nanotechnology Division, National Institute of Standards and Technology, Gaithersburg, USA}
\date{\today}


\title{Two-Dimensional Nonlinear Mixing Between a Dissipative Kerr Soliton and Continuous Waves for a Higher-Dimension Frequency Comb}

\begin{abstract}
   Dissipative Kerr solitons (DKSs) intrinsically exhibit two degrees of freedom through their group and phase rotation velocity. Periodic extraction of the DKS into a waveguide produces a pulse train and yields the resulting optical frequency comb's repetition rate and carrier-envelope offset, respectively. Here, we demonstrate that it is possible to create a system with a single repetition rate but two different phase velocities by employing dual driving forces. By recasting these phase velocities into frequencies, we demonstrate, experimentally and theoretically, that they can mix and create new phase-velocity light following any four-wave mixing process, including both degenerately pumped and non-degenerately pumped effects. In particular, we show that a multiple-pumped DKS may generate a two-dimensional frequency comb, where cascaded nonlinear mixing occurs in the phase velocity dimension as well as the conventional mode number dimension, and where the repetition rate in each dimension differs by orders of magnitude.
 \end{abstract}

\maketitle


Timekeeping~\cite{DiddamsScience2001, PappOpticaOPTICA2014a, NewmanOptica2019}, ranging~\cite{CaldwellNature2022, RiemensbergerNature2020}, astronomical instrument calibration~\cite{MetcalfOpticaOPTICA2019, MurphyMonthlyNoticesoftheRoyalAstronomicalSociety2007}, and coherent microwave-optical links~\cite{DelHayeNaturePhoton2016, SpencerNature2018} are just a few examples of the many metrology-related applications that have benefited greatly from the development of optical frequency combs (OFCs). OFCs can be generated by various techniques involving electro-optic modulation or mode-locking of a pulsed laser. Utilizing the Kerr nonlinearity of a resonator to create a cavity soliton, also known as a dissipative Kerr soliton (DKS), is another increasingly appealing approach. Since their initial experimental fiber demonstration~\cite{LeoNaturePhoton2010b}, DKSs have become the preferred method for OFCs on-chip~\cite{HerrNaturePhoton2014,BraschScience2016, LiOptica2017a}, where a intracavity single DKS state is extracted periodically to create a fixed pulse train. This pulse train is the foundation for the OFC through the Fourier relationship. The advent of high-quality microresonators employing foundry-like manufacturing processes based on highly nonlinear materials~\cite{JungOpt.Lett.OL2013, WangNatCommun2019a, BraschScience2016, LiOptica2017a} has made it possible for OFCs to be used in field-deployable applications, owing to their compact footprint and low-power CW excitation~\cite{SternNature2018a, MoilleLasersPhotonicsRev.2020a}. The carrier-envelope offset (CEO), appearing as a frequency shift with respect to DC of the comb in the frequency domain or as a fast oscillation phase offset from the envelope in the time domain, is an essential property of OFCs~\cite{PicqueNaturePhoton2019,SpencerNature2018, DrakePhys.Rev.X2019}. In the case of DKSs, this CEO arises from the fundamental difference between the phase and group velocities of the soliton within the resonator. Periodic DKS extraction by the coupling bus at every round trip results in a constant increase of the offset between the carrier envelope and the rapid oscillation for each pulse component. Consequently, fully locking a DKS-based OFC demands both repetition rate and CEO frequency locking~\cite{NewmanOptica2019, SpencerNature2018, BraschLightSciAppl2017}, essentially locking the two degrees of freedom of the soliton: its group and phase velocities. One significant avenue of research on integrated OFCs has been on optimizing the group velocity dispersion by employing approaches like finely tailoring photonic mode confinement~\cite{okawachi_bandwidth_2014, MoilleAppl.Phys.Lett.2021a} or exploiting unique photonics characteristics~\cite{KimNat.Commun.2017, MoilleOpt.Lett.OL2018, Moille2022d} to meet milestones otherwise only possible with fiber-based spectral broadening of a mode-locked laser, such as broadband, octave-spanning bandwidths~\cite{OkawachiOpt.Lett.OL2011a, LiOptica2017a, YuPhys.Rev.Applied2019a}. To the extent to which the CEO is considered in most DKS experiments, it is primarily as a key parameter that must be fixed and locked to provide metrological stability~\cite{NewmanOptica2019, SpencerNature2018, BraschLightSciAppl2017}. Recently, the study of DKSs subjected to various driving forces has garnered increased attention, thanks to their fundamental~\cite{ZhangPhys.Rev.Lett.2022, TaheriNatCommun2022} and applied~\cite{ZhangNatCommun2020, MoilleNat.Commun.2021a} novelties. Athough they have the capacity to generate record-breaking frequency spans~\cite{MoilleNat.Commun.2021a}, multi-pumped DKS-OFCs consist of composite comb components at the same repetition rate, yet frequency shifted relative to one another~\cite{MoilleNat.Commun.2021a, QureshiCommunPhys2022}. This presents a challenge for metrology applications since this frequency interleaving introduces a new degree of freedom that must be locked. Thus, while multiply-pumped microresonators hosting DKS combs have been used for spectral translation~\cite{MoilleNat.Commun.2021a}, The origin of this constant frequency offset has not been thoroughly investigated, nor has it been studied if this degree of freedom may be utilized to create novel nonlinear optical processes.\\
\begin{figure*}[t]
    \begin{center}
        \includegraphics[width = 0.95\textwidth]{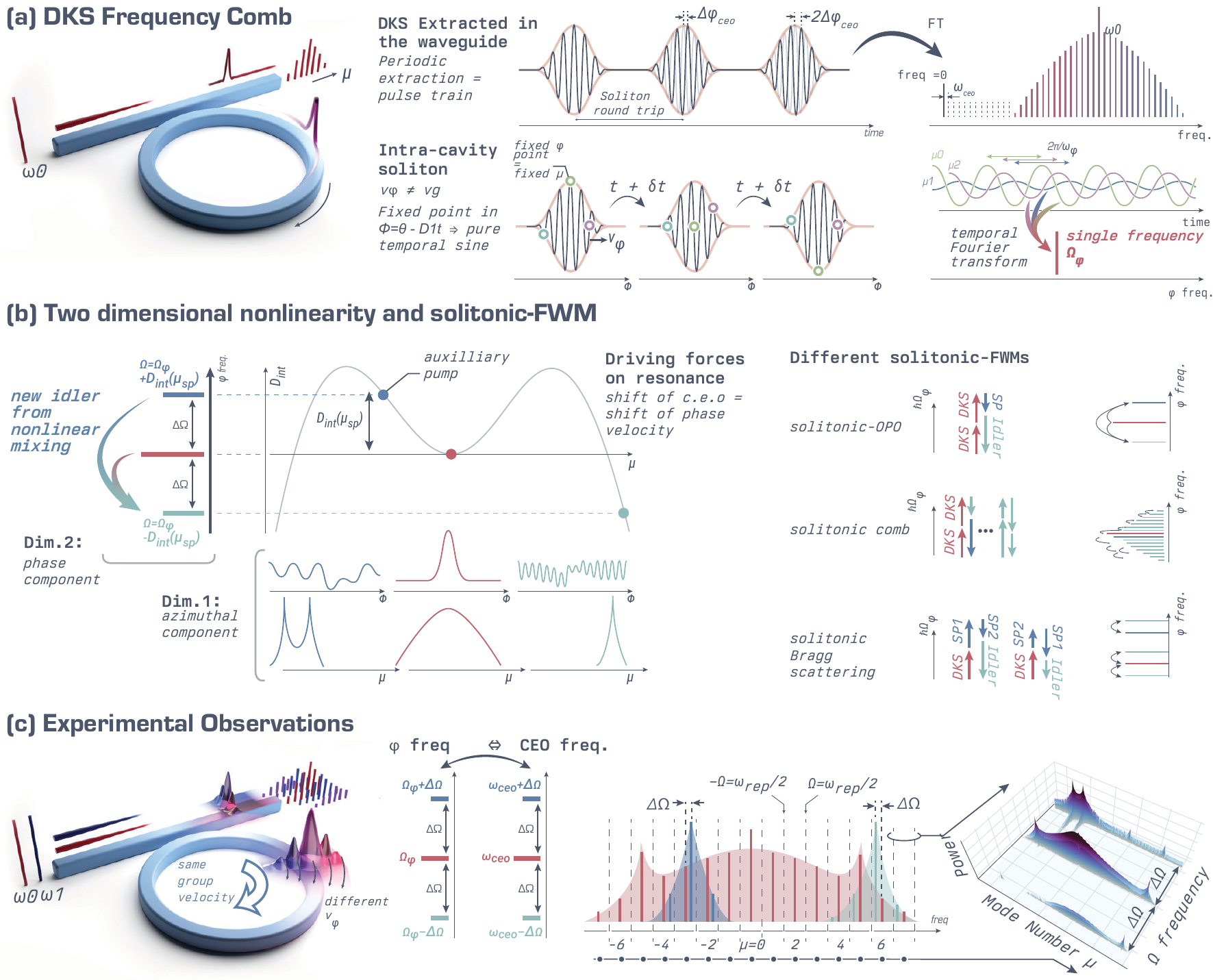}
    \end{center}
    \caption{\label{fig:1} \textbf{Principle of a two-dimensional frequency comb from harnessing group and phase velocity} -- %
    \textbf{(a) Standard DKS frequency comb.} The CEO in the pulse train is produced by periodic extraction at the waveguide coupling point and the mismatch in the DKS phase and group velocity. Different  points within the DKS waveform will all oscillate at the same frequency, determined by the rapid oscillation drift from the envelope at any point in its moving frame $\phi$. The difference between group velocity and phase velocity causes this drift. A single DKS is, therefore, equivalent to a single frequency $\Omega_\varphi$ in its moving frame. %
    \textbf{(b) Phase velocity mixing.} Thanks to the cavity dispersion (central plot of the integrated dispersion $D_\mathrm{int}$), another phase velocity may be introduced when an auxiliary pump laser is applied, as the auxiliary pump needs to be on resonance for effective energy transfer to the ring. The CEO frequency shift from the DKS comb is directly tied to the integrated dispersion, therefore yielding a shift in the phase-velocity frequency of $\Delta\Omega\simeq D_\mathrm{int}(\mu_\mathrm{sp})$. The ``colors'' at these two separate frequencies ($\Omega_{\varphi}$ and $\Omega_{\varphi}+\Delta\Omega$) can combine in a nonlinear medium to produce another ``color'' at a third (idler) frequency at $\Omega_{\varphi}-\Delta\Omega$ (left image). These phase velocity frequencies each have spectral components that populate many modes in the azimuthal direction along the ring (bottom left image). As a result, a two-dimensional system is produced. This specific nonlinear mixing process is one among others possible: optical parametric oscillation, cascaded parametric oscillation to form a two-dimensional comb, and four-wave mixing Bragg scattering. %
    \textbf{(c): Experimental observables}. In our experiments, the intracavity field is not directly probed, and only the periodically-extracted ``colors'' coupled into the waveguide are observed (left image). The different phase velocity frequencies correspond to wavepacket components at the same repetition rate (group velocity) yet different phase velocities. Therefore, these extracted ``colors'' in the waveguide will experience different CEO frequencies. As a result, the generated frequency comb will consist of several components, each offset by $\pm\Delta\Omega$ from the DKS CEO frequency. The obtained frequency comb can be transformed into a three-dimensional plot (right image) where the $z$ axis is the comb tooth power, the $x$ axis is $\mu$, and the $y$ axis is $\Omega$, \and helps emphasize its two-dimensional nature. This is done by assuming that each mode $\mu$ is composed of any power within the frequency range $\Omega=[-\omega_\mathrm{rep}/2;  \omega_\mathrm{rep}/2]$ around a DKS comb tooth. Here, only the soliton-OPO, hence three comb components in $\Omega$, is represented for simplicity. The solitonic-comb where this process cascades into many more teeth in $\Omega$ is presented in \cref{fig:4}, while solitonic-FWM-BS is demonstrate in  \cref{fig:5}.
    }
\end{figure*}%
\indent In this work, we start by considering how the difference between the phase and group rotation velocities of a DKS may be interpreted as a single intrinsic DKS frequency. When injecting a secondary driving pump, the dispersion of the resonator results in an intracavity field component with a distinct phase rotation velocity from the DKS, and thus another single phase-velocity frequency which can mix with the soliton. This creates a new ``color'' of the DKS, and in some circumstances, can result in a cascading effect that produces a two-dimensional frequency comb. This two-dimensional comb is characterized by two distinct repetition rates, one being the conventional DKS repetition rate representing the frequency spacing between spectral components that populate adjacent azimuthal modes of the resonator, and the second being a repetition rate in the phase-velocity frequency space. We start our investigation by outlining the theoretical background necessary to comprehend the two dimensions involved in the two-dimensional frequency comb. We then empirically confirm that the interplay between the dispersion-less nature of the DKS and the cavity dispersion leads to a shift of phase velocity relative to the DKS at the secondary pump, which shows a shift of the CEO. Then, we demonstrate this novel nonlinear mixing by showing an OPO in the phase velocity space, and that this FWM process may cascade into a 2D frequency comb. We characterize this 2D comb through an optical measurement, which we show is consistent with a full LLE model result. We highlight that the repetition rate in the phase rotation velocity domain is constant, producing a radio frequency (RF) comb that is three orders of magnitude lower than the THz repetition rate in the azimuthal mode ($\mu$) dimension. Finally, we demonstrate non-degenerate four-wave mixing Bragg scattering within the phase velocity space, illustrating the generality with which DKS nonlinear mixing can occur in this second dimension.

\vspace{1ex}
\textbf{Physical description of a bichromatically-pumped DKS --} %
First, we aim to remind the reader about the difference in group and phase velocity of a DKS and how it can be recast as a single frequency (Fig.~\ref{fig:1}(a)). The DKS can be expressed in its moving frame as  $E(\phi, t) = \sqrt{2A_0} \sech(\sqrt{A_0/d_2}\phi)\exp\left[i(\Omega_\varphi t + \phi_0) \right] + E_0$, where $\phi = \theta - \omega_\mathrm{rep}t$, $\theta$ is the azimuthal angle of the resonator, $d_2$ describe the group velocity dispersion, $\omega_\mathrm{rep}=v_\mathrm{g}/R$ is the group velocity repetition rate across the cavity for radius $R$ (\textit{i.e.} the angular frequency for the DKS to come back to the same $\theta$ point), $\Omega_\varphi = \frac{v_\varphi - v_g}{R}$ is the group $v_g$ and phase $v_\varphi$ velocity frequency mismatch, and $E_0$ is the background on which the DKS sits on. Periodic extraction of the pulse yields the carrier-envelope offset $\Delta\varphi_\mathrm{ceo} = 2\pi \Omega_\varphi /\omega_\mathrm{rep} \;(\bmod\; 2n\pi \;n\in\mathbb{N})$ and thus the carrier-envelope offset observed in the pulse train is similar -- albeit modulo $2\pi$ -- to the phase velocity frequency $n \omega_\mathrm{ceo} \approxeq \Omega_\varphi $. Consequently, for each point in the $\phi$ coordinate traveling with the DKS, a single frequency $\Omega_\varphi$ may be measured experimentally through the carrier envelope offset. Interestingly, it can also be observed in the Fourier domain given all azimuthal Fourier components $\mu$ (relative to the pumped mode) from $\phi$, which once extracted yield the comb teeth, carry $\omega_\mathrm{ceo}$.

Now, let us assume that the resonator is being pumped bichromatically at $\omega_0$ and $\omega_\mathrm{sp}$ with a single DKS generated by the primary pump. Through cross-phase modulation, the intra-cavity wave generated by the secondary pump at $\mu_\mathrm{sp}$ locks its repetition rate to the DKS~\cite{WangOptica2017,MoilleNat.Commun.2021a}. Hence, the transformation $\theta \mapsto \phi$ still holds. Similarly, any point in $\phi$ will carry the same frequency $\Omega_\varphi$ from the DKS. Due to the cavity dispersion, the second ``color" of the wavepacket is created at a different phase velocity by the secondary pump, which must be in resonance for light injection into the cavity. The integrated dispersion $D_\mathrm{int}(\mu) = \omega_\mathrm{res} - (\omega_0 + \mu\omega_\mathrm{rep})$, beyond describing the dispersion of the resonator, also represents the shift of CEO frequency of the resonances $\omega_\mathrm{res}$ from the fixed frequency markers $\omega_0 + \mu\omega_\mathrm{rep}$ being the output measured frequency comb. The carrier envelope offset and phase rotation velocity are directly related, as was previously shown. As a result, the secondary pump wavepacket ``color" displays a change in phase velocity frequency with respect to the DKS. When the frequency detuning of the secondary pump $\delta\omega_\mathrm{sp}$ is accounted for, the bichromatically-pumped DKS translates into a system with two distinct phase velocities separated by $\Delta\varphi = 2\pi \Delta\Omega/\omega_\mathrm{rep}$, where $\Delta\Omega = \Omega_\mathrm{sp} - \Omega_\varphi = D_\mathrm{int}(\mu_\mathrm{sp}) + \delta\omega_\mathrm{sp}$ is the shift between the two single distinct phase-velocity frequencies for each ``color'' of the system [\cref{fig:1}(b)]. Being in a third-order nonlinear material — an inherent property for the existence of a DKS — these frequencies can mix following an OPO-like interaction, producing a new frequency $\Omega_\mathrm{idl} = 2\Omega_{\varphi} - \Delta\Omega$. This \textit{solitonic}-OPO is only an illustration of the potential nonlinearity between CW light and a DKS that may exist in the phase-velocity space. However, all other FWM processes are theoretically possible, like cascading the solitonic-OPO into a comb in the phase velocity space and using noiseless frequency conversion like FWM Bragg scattering. This new phase velocity frequency corresponds to a $-\Delta\Omega$ offset of the carrier-envelope from the DKS, which is observable in the outcoupled frequency comb [\cref{fig:1}(c)]. The phase matching with the cavity resonance will indicate the location of the resonant enhancement of this third wave in the $\mu$ domain.

\vspace{1ex}
\textbf{Theoretical model for the two-dimensional frequency comb --} %
Frequency comb modeling and simulations may adhere to either the coupled mode theory (CMT) or the Lugiato-Lefever equation (LLE) for frequency or time domain analysis, respectively~\cite{chemboModalExpansionApproach2010, ChemboPhys.Rev.A2013}. The Fourier transform connects these two models, which are fundamentally comparable. For our multi-pumped system, one may use an equivalent to the CMT formalism to describe the nonlinear mixing processes discussed in the previous section, in which the phase-velocity frequencies are mixing into what we have termed a ``\textit{solitonic-OPO}" to produce new colors in this frequency domain. Yet, as we've established before, these new frequency components are also wavepacket colors in the $\mu$ — subsequently the $\phi$ — domain. In lieu of modal CMT, thus, the system can be described by a set of coupled LLEs:

\begin{align}
    \partial_t a_\sigma(\phi, t) &= \left(-\frac{\kappa}{2} - \sigma \Delta\Omega \right)a_\sigma + i\sum_\mu D_\mathrm{int}(\mu)A_\sigma(\mu, t)\mathrm{e}^{i\phi\mu} \nonumber\\
    &+ i\gamma L\sum_{\alpha, \beta}a_\alpha a_\beta^*a_{\alpha-\beta + \sigma} \nonumber\\
    &+ \delta_0 \sqrt{\kappa_\mathrm{ext}}F_0\mathrm{e}^{(i\delta\omega_0 t)} 
    + \delta_1 \sqrt{\kappa_\mathrm{ext}}F_1\mathrm{e}^{i(\delta\omega_1 t + \mu_\mathrm{sp}\phi)} 
    \label{eq:1}
\end{align}

\noindent with $a_\sigma(\phi, t)$ the temporal wavepacket ``color'' of the phase velocity frequency comb at the index $\sigma$, the total loss rate is $\kappa$ with the external loss rate (\textit{i.e.} coupling \ks{rate}) $\kappa_\mathrm{ext}$, $\Delta\Omega$ the repetition rate of the comb in $\sigma$, $A_\sigma(\mu, t) = \mathrm{FT}_\mu \left[ a_\sigma(\phi, t) \right]$ the azimuthal Fourier transform of the wavepacket color, $F_j$ is the driving force at the $j$\textsuperscript{th} color, $\delta_j=1$ if $\sigma = j$ and 0 otherwise, $\alpha, \beta$ are the other wavepacket colors (\textit{i.e.} $\sigma$ ``colors'') that are involved in nonlinear mixing following FWM phase matching. This term is similar to modal four-wave mixing in the CMT, which describes conventional microresonator frequency combs. In the case of only three waves considered, similar to the example described previously -- the DKS is $\sigma=0$, the secondary pumped color is $\sigma=1$, and the new idler color is $\sigma=-1$. It is also worth noting the absence of dispersion in the $\sigma$ dimension following the assumption of small $D_\mathrm{int}\ll\omega_\mathrm{rep}$ and $\max(\partial_\mu D_\mathrm{int})<\kappa$ in the modal region of interest. To confirm the suggested model, we will compare the experimental demonstration with a multi-pump LLE~\cite{TaheriEur.Phys.J.D2017} with the following form, from which we may get the various CEOs:

\begin{align}
    \partial_t a(\phi, t) &=  -\frac{\kappa}{2} a + i\sum_\mu D_\mathrm{int}(\mu)A(\mu, t)\mathrm{e}^{i\phi\mu} +i\gamma L|a|^2a\nonumber\\
    &+\sqrt{\kappa_\mathrm{ext}}F_0\mathrm{e}^{(i\delta\omega_0 t)} 
    + \sqrt{\kappa_\mathrm{ext}}F_1\mathrm{e}^{i(\Delta\Omega t + \mu_\mathrm{sp}\phi)} 
    \label{eq:2}
\end{align}

The simulation protocol, which we cover in greater detail in the methods, entails closely replicating the experiment using the open-source \textit{pyLLE} software~\cite{MoilleJ.RES.NATL.INST.STAN.2019}. The resultant simulated DKS is retrieved at precisely each round-trip, reproducing the pulse train in the output waveguide, which may thereafter be Fourier transformed. Unlike the Fourier transform of the azimuthal profile of the DKS within the cavity, which only allows one to obtain the comb envelope profile, the frequency comb from the pulse train allows one to resolve the different CEOs and, as a result, retrieve the various phase velocity offsets. Given that this model accounts for all FWM processes, the coupled-LLE model will be validated if it can duplicate both experimental results and does not display any frequency comb components other than at the CEO frequencies given by $\sigma\Delta\Omega$.

\begin{figure}[htpb]
    \begin{center}
        \includegraphics[width = \columnwidth]{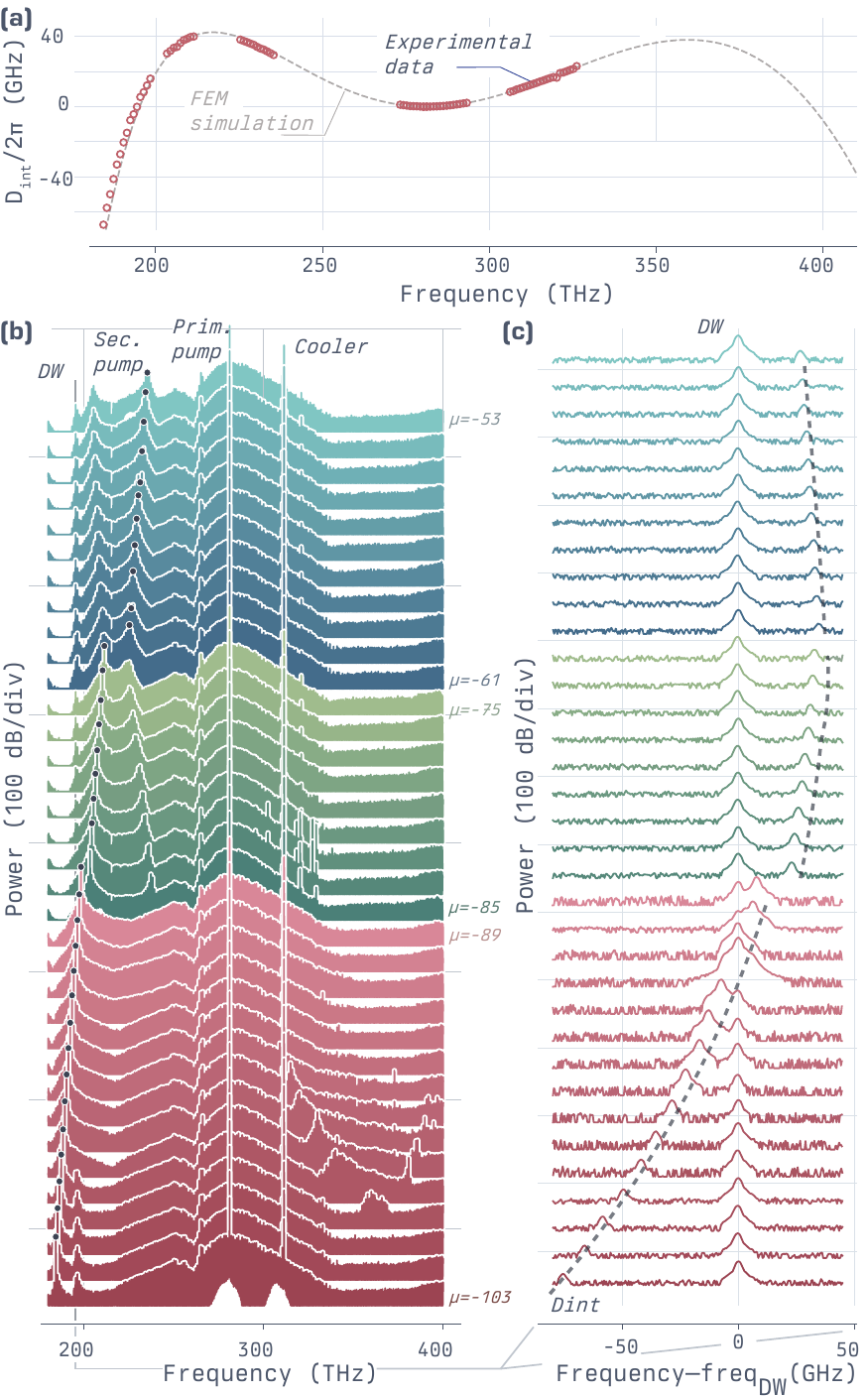}
        \caption{\label{fig:2}\textbf{Carrier envelope offset shift from cavity dispersion -- }%
        \textbf{(a)} Integrated dispersion of the microresonator, with the experimental measurements in red and the FEM simulations in dashed grey. %
        \textbf{(b)} DKS frequency comb from a 283~THz primary pump, with different auxiliary pump (black circle) for $\mu=-53$ to $\mu=-61$ (teal), $\mu=-75$ to $\mu=-85$ (green), and $\mu=-89$ to $\mu=-103$ (red). %
        \textbf{(c)} Zoom-in of the frequency combs around the DKS frequency comb dispersive wave (DW) at $\mu=-92$ for the different auxiliary pump, with $D_\mathrm{int}(\mu)$ shown in dashed-grey line matching the comb tooth offset of the auxiliary, pumped frequency comb component.%
        }
    \end{center}
\end{figure}

\begin{figure*}
    \begin{center}
        \includegraphics[width = \textwidth]{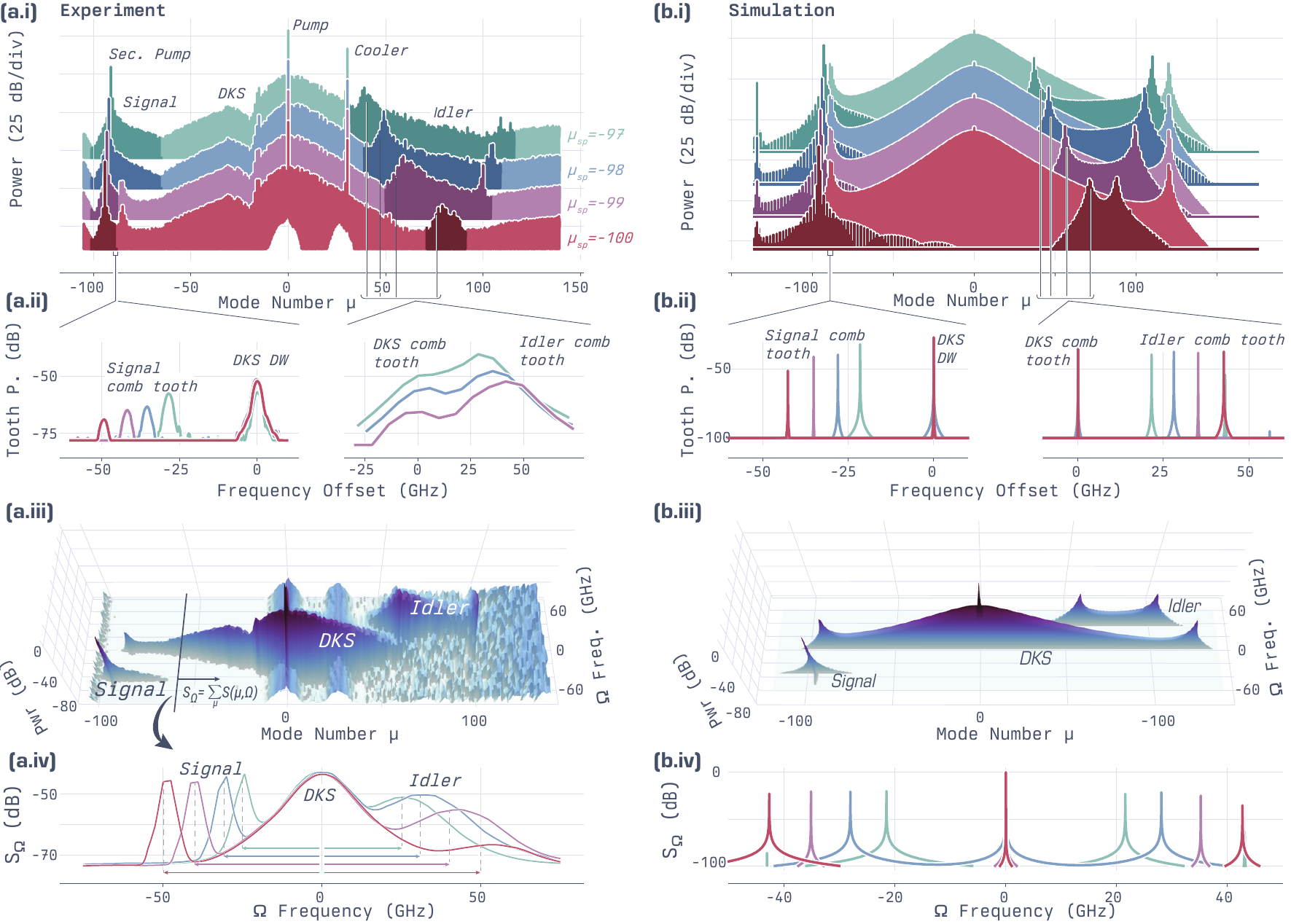}
    \end{center}
    \caption{\label{fig:3} \textbf{Solitonic-optical parametric oscillation -- }%
    \textbf{(a.i)} Experimental OFC obtained from primary pumping at 283~THz with auxiliary pumping at $\mu_\mathrm{sp}=-97, -98, -99, -100$. The different frequency comb components that are frequency offset from the DKS one are highlighted. %
    \textbf{(a.ii)} Zoom onto the frequencies of interest. (Left) Zoom-in around the dispersive wave of the DKS comb component, similar to~\cref{fig:2}(c). (Right) Zoom-in where the idler comb component power is large, showing the equal but opposite sign frequency shift. %
    \textbf{(a.iii)} Projection of the OFC auxilliary pumped at $\mu_\mathrm{sp}=-99$ into the $\{\mu, \Omega\}$ space such that, for a given $\mu$, each component in $\Omega$ corresponds to the power in an interval of $[-\omega_\mathrm{rep}/2; +\omega_\mathrm{rep}/2]$ around each $\omega_\mu$. %
    \textbf{(a.iv)} Integration over $\mu$, highlighting the equal and opposite sign of the CEO frequency shift, hence the phase velocity frequency, and following the OPO energy conservation requirements. %
    \textbf{(b.i-iv)} LLE simulations following the protocol to reproduce the experiment. The agreement with the experimental observation is clear, where three distinct comb components exist, following the solitonic-OPO energy conservation in the CEO, and thus the phase velocity. Interestingly, no other nonlinear mixing is present, which demonstrates the validity of the coupled-LLE model in~\cref{eq:1}. %
    }
\end{figure*}

\vspace{1ex}\textbf{Introducing another phase velocity in the system:} The basic principle that we have put forth is based upon the discrepancy between the DKS's dispersion-less nature and the resonator's dispersion, which creates a shift in CEO and, as a result, a shift in phase velocity, enabling the nonlinear mixing process. We use a $H=670$~nm thick, $R=23$~{\textmu}m radius, $RW=830$~nm width \ce{Si3N4} microring resonator embedded in \ce{SiO2}, which yields $\omega_\mathrm{rep}/2\pi\simeq 997$~GHz. As described previously, the integrated dispersion $D_\mathrm{int}$ shall drive the CEO shift in the bichromatically pumped system. The resonator dispersion of the fundamental transverse electric mode is experimentally determined [\cref{fig:2}(a)] using the various available continuously tunable lasers (CTL) with the resonance frequencies calibrated using a wavemeter. Accurate reproduction of this measurement is achieved by dispersion simulation utilizing the finite element method. The only limitation on the modes $\mu$ that may be studied, either linearly or nonlinearly, are the CTL tuning range availability. We then generate a single DKS pumped with 180~mW on-chip at 283~THz while using a 306~THz counterpropagating and cross-polarized cooler pump to access the DKS state adiabatically. We use an auxiliary pump on resonance at various modes $\mu$ [\cref{fig:2}(b)] to test our hypothesis and explore the frequency shift between the comb components. One can resolve this CEO frequency shift [\cref{fig:2}(c)] using an optical spectrum analyzer (OSA) with a resolution of 20~pm (2.5~GHz at 1550~nm). The previously measured $D_\mathrm{int}$ at the auxiliary pump frequency closely matches the obtained shift in CEO frequency between the DKS comb and the auxiliary pumped component. We are able to accurately measure this by looking at the frequency difference between the auxiliary pumped component and the dispersive wave (DW) in the DKS comb, the latter being fixed by $D_\mathrm{int}(\mu)$. Through this measurement, we validate the first hypothesis of this study, which states that the phase velocity frequency offset is driven by $D_\mathrm{int}(\mu_\mathrm{sp})$.

\vspace{1ex} \textbf{Nonlinear mixing in the phase velocity domain: solitonic-OPO: } Now that we have demonstrated the ability to introduce a phase velocity shift in the system by simply auxiliary pumping one resonance at a mode $\mu_\mathrm{sp}$, which introduces a CEO frequency shift of $\Delta\Omega= D_\mathrm{int}(\mu_\mathrm{sp}) + \delta\omega_\mathrm{sp}$, we demonstrate nonlinear mixing in this phase-velocity space that is orthogonal to the azimuthal mode space $\mu$. Let us first examine the most basic FWM scenario, which corresponds to the solitonic-OPO situation. We employ the same microring resonator previously investigated. We optically pump the system with a 185~mW on-chip primary pump at 283~THz and auxiliary pump it with 10~mW of on-chip power at modes $\mu_\mathrm{sp}=-97, -98, -99$ and, $-100$, which corresponds to $\omega_\mathrm{sp}/2\pi  = 189, 188, 187$ and, $188$~THz, respectively. The generated comb spectra are shown in Fig.~\ref{fig:3}(a,i). Aside from the formation of a comb component around the auxiliary pump, another wavepacket color is created at a higher frequency with DW-like at modes $\mu_\mathrm{idl}$ that should meet the phase matching such that $\Omega_\mathrm{sp} -\Omega_\varphi = \Omega_\varphi - \Omega_\mathrm{idl}$, as predicted by the theory. One may zoom closely on the relevant comb teeth to confirm this behavior, as shown in Fig.~\ref{fig:3}(a,ii). As predicted by the previously stated CEO frequency shift, the auxiliary pump component acting as the signal in the phase-velocity frequency space $\Omega$ exhibits a shift from the DKS DW of $\Delta\Omega$, which varies with the auxiliary pumped mode according to the integrated dispersion. On the other hand, the idler wavepacket color exhibits the opposite sign of CEO frequency shift. It is worth noting that the sharpness of the comb teeth is restricted by the OSA's 20~pm resolution, which corresponds to around 6~GHz at 980~nm (thus a threefold decrease in resolution when compared to the 1550~nm spectrum). As indicated in the prior physical description of the process section, it is illuminating to recast the obtained frequency comb into a two-dimensional coordinate system, including the mode number and the CEO frequency shift $S(\mu, \Omega)$. For a given comb tooth $\mu$ (the $x$ axis), the power inside $\Omega = [-\omega_\mathrm{rep}/2; +\omega_\mathrm{rep}/2]$ corresponds to the new system's $y$ axis. Such transformation underlines the multi-component nature of the frequency comb, as illustrated in \cref{fig:3}(a.iii) or one specific secondary pump frequency, where the signal (\textit{i.e.}, the auxiliary pumped component) is equally distant from the DKS comb as the idler component. One may go further by integrating the CEO frequency shift with the mode number so that $S_\mathrm{\Omega} = \sum_\mu S(\mu, \Omega)$, indicating that each component displays a single CEO frequency shift evenly spaced from the DKS (Fig.~\ref{fig:3}(a,iv)). This single CEO shift highlights the offset in only the phase velocity space while obeying the fundamental OPO energy conservation requirement, demonstrating the \textit{solitonic}-OPO harnessing nonlinearity in an orthogonal dimension from the azimuthal modes.

\begin{figure*}[t]
    \begin{center}
        \includegraphics[width = \textwidth]{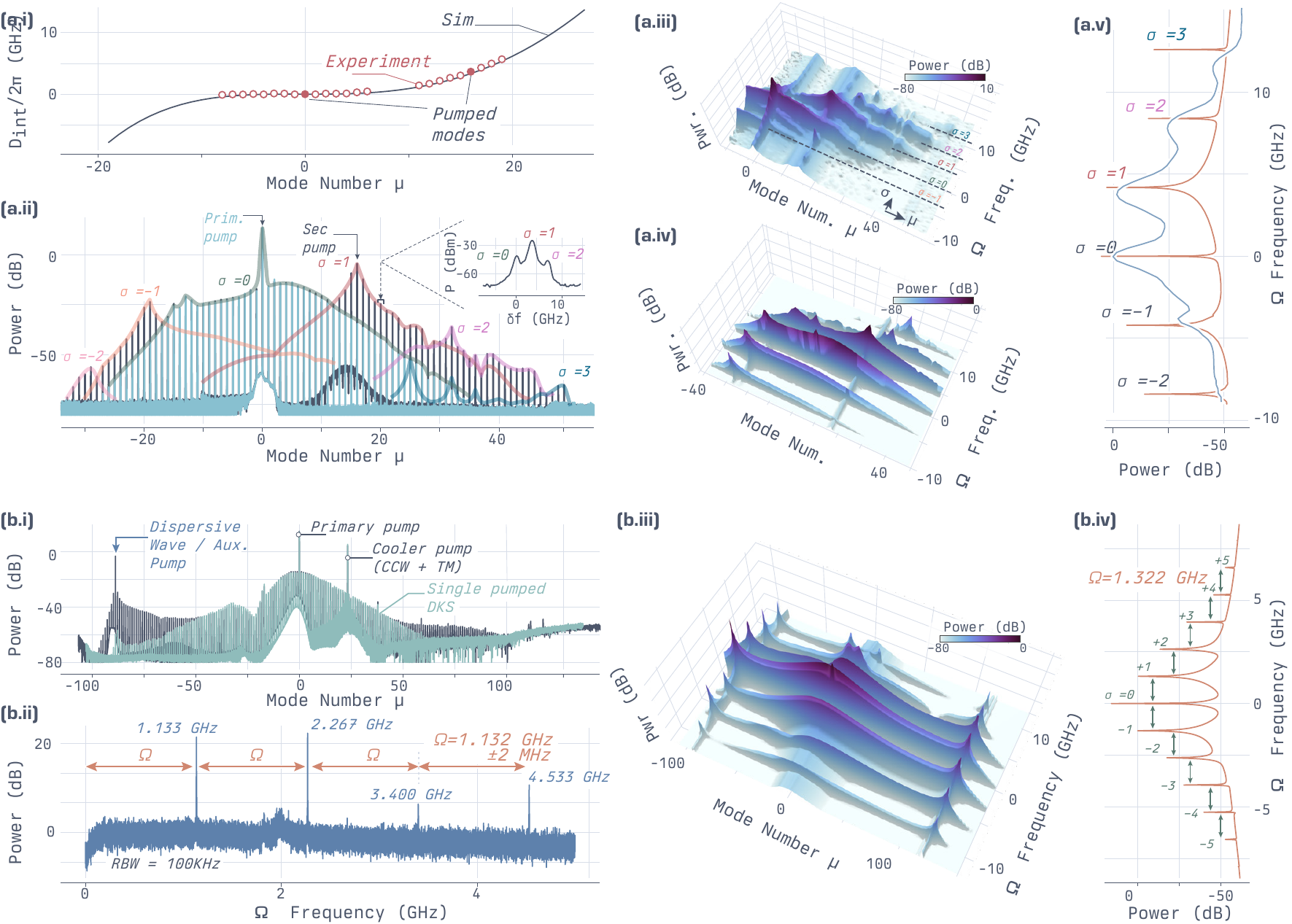}
    \end{center}
    \caption{\label{fig:4} \textbf{Two dimensional frequency comb from cascaded solitonic-OPO} -- %
    \textbf{(a.i)} Measured (circles) and simulated (line) integrated dispersion of the microring resonator used. The pumped modes are highlighted with the auxiliary pump at $\mu=16$. %
    \textbf{(a.ii)} Experimental OFC under single (teal) and dual pumping (dark blue). Under dual pumping, multiple idlers can be seen with multiple corresponding frequency offsets from the DKS comb. %
    \textbf{(a.iii)} Projection of the comb into the $\{\mu, \Omega\}$ space highlighting the cascading solitonic-OPO nonlinearity. %
    \textbf{(a.iv)} LLE simulation reproducing the experimental cascaded solitonic-OPO accurately. %
    \textbf{(a.v)} Integration of the two-dimensional comb along the $\mu$ domain in experiment (blue) and simulation (red), highlighting the agreement of the tooth spacing in the $\sigma$ domain. %
    \textbf{(b.i)}  Experimental OFC under single (teal) and dual pumping (dark blue) for the microring used in \cref{fig:2,fig:3} when auxiliary pumping at the DKS DW. %
    \textbf{(b.ii)}  Electrical spectrum of the dual-pump microcomb in the $\sigma$ domain exhibiting a set of 4 comb teeth frequency spaced by the same repetition rate $\Delta\Omega$. %
    \textbf{(b.iii)} From LLE simulation, one could resolve the two-dimensional frequency comb in $\{\mu, \Omega\}$ that corresponds to the experiment. 
    \textbf{(b.iv)} The 2D-comb exhibits a set of 11 $\sigma$ comb teeth that are highlighted through the $\mu$ integration of the 2D comb $S_\Omega$. %
    }
\end{figure*}

Although the solitonic-OPO viewpoint and the model described in \cref{eq:1} are in qualitative agreement with these experimental results, a comprehensive simulation employing the single multi-pump LLE from \cref{eq:2} that recreates the experiment would provide a deeper understanding of the nonlinearities in play. Following the simulation protocol described in the methods, the frequency comb can be accurately reproduced (Fig.~\ref{fig:3}(b)) using the resonator dispersion from Fig.~\ref{fig:2}(a), on-chip power from the experiment, the nonlinearity coefficient $\gamma=3.2$~W\textsuperscript{-1}$\cdot$m\textsuperscript{-1} from FEM simulation, and losses obtained from linear measurement $\kappa = 2\kappa_\mathrm{ext} \simeq 200$~MHz. One may observe the system's three distinct components: the signal, idler, and DKS. Similarly, the signal and idler exhibit an equal but opposing frequency shift in CEO (Fig.~\ref{fig:3}(b,ii)), which reveals similar behavior to the experiment when recasting in the $\{\mu, \Omega\}$ space (Fig.~\ref{fig:3}(b,iii)). Finally, only the three anticipated tones in $\Omega$ are displayed by the integration along $\mu$ (Fig.~\ref{fig:3}(b,iv)). It is fair to note that only the solitonic-OPO is present, which validates the coupled-LLE model in \cref{eq:1} that we have proposed, given that \cref{eq:2} account for all potential FWM processes.

\begin{figure*}
    \begin{center}
        \includegraphics{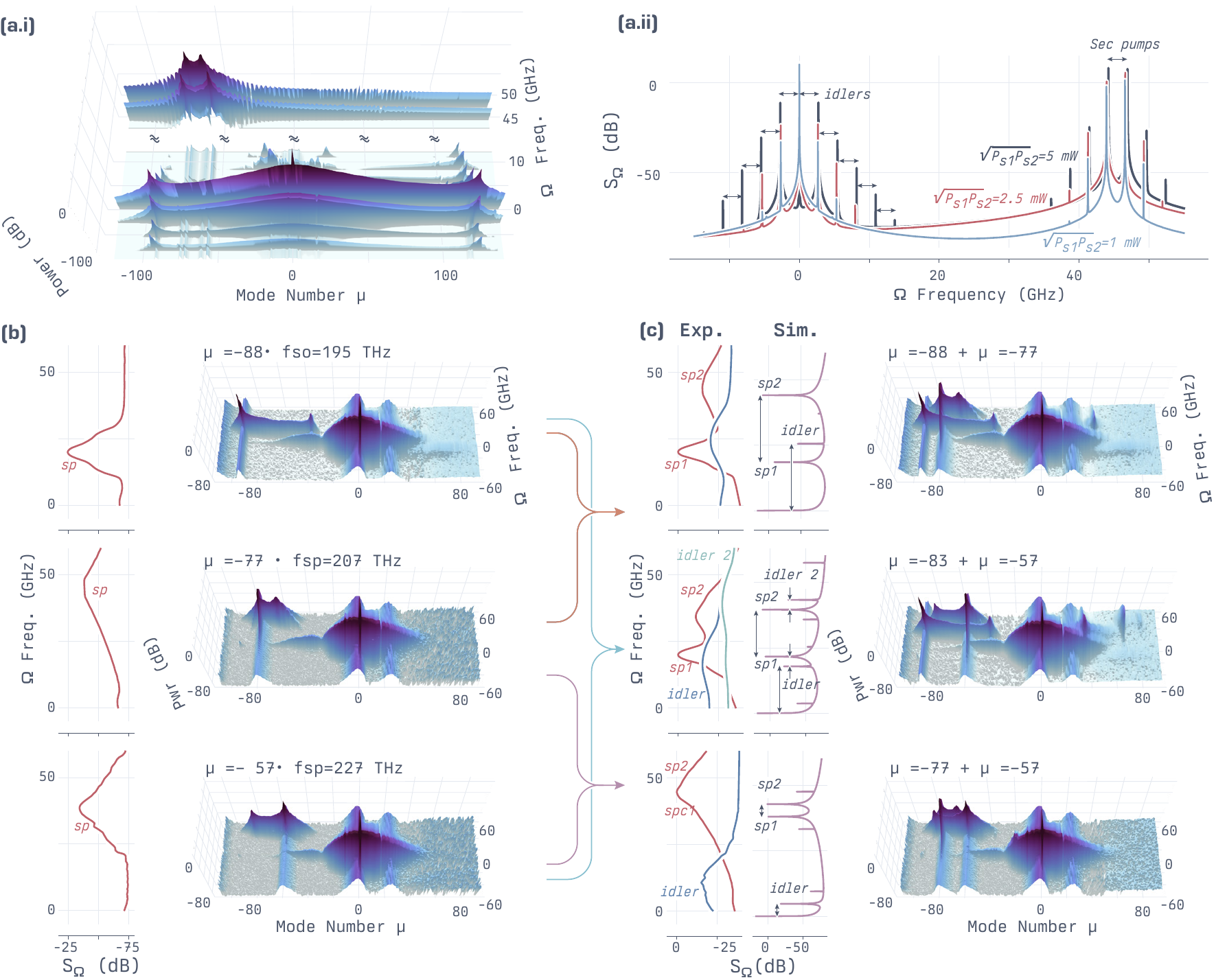}
    \end{center}
    \caption{\label{fig:5} \textbf{FWM Bragg scattering in the phase  velocity -- } %
    \textbf{(a.i)} LLE simulation with two auxiliary pumps that produce a scattering of the DKS in the $\Omega$ phase-velocity frequency domain. %
    \textbf{(a.ii)} Simulated $S_\Omega$ using two auxiliary pumps at different power shows the possibility for FWM-BS of the DKS, which can also further cascade, creating a synthetic frequency lattice. %
    \textbf{(b)} Experimental measurement of $S_\Omega$ (left) and the corresponding comb in the $\{\mu, \Omega\}$ space for three different auxiliary pump positions. Only one tone in the $\Omega$ frequency space can be seen (the idler is not out-coupled). %
    \textbf{(c)} Experimental data from the three different combinations of dual-auxiliary pumping positions from \textbf{(b)}, including $S_\Omega$ (left) and the corresponding spectrum in the $\{\mu, \Omega\}$ space (far right). For each of the combinations, a new idler is created at $\Omega = |\Omega_\mathrm{sp1} - \Omega_\mathrm{sp2}|$, in excellent agreement with the LLE simulations (left and center).
    } 
\end{figure*}%

\vspace{1ex}
\textbf{Two dimensional frequency comb -- } No restriction prevents the phase velocity frequencies from cascading further, similar to the typical creation of modulation instability comb and cavity solitons in the $\mu$ domain, in which cavity mode photons mix first to produce an OPO and then cascade into a comb. Additional colors can appear from the three-color wavepacket produced by phase velocity frequency mixing shown above, each separated by the same frequency $\Delta\Omega$ such that $\Omega (\sigma) = \Omega_\varphi + \sigma\Delta\Omega$. This produces an evenly spaced frequency comb in the phase rotation velocity frequency $\Omega$ space. Each comb tooth, indexed by $\sigma$, has an azimuthal ($\mu$ domain) envelope. As a result, every $\sigma$ comb tooth is also in $\mu$, a frequency comb. The resultant system is an orthogonal two-dimensional frequency comb defined by $\mu$ and $\sigma$ with the characteristic frequencies $\omega_\mathrm{rep}$ and $\Delta\Omega$ with $\Delta\Omega \ll \omega_\mathrm{rep}$. In the experiment, this 2D-frequency comb is characterized by a set of comb teeth in $\mu$ separated by $\omega_\mathrm{rep}$, and a set of comb teeth in $\sigma$ surrounding each mode $\mu$ and spectrally located within $[-\omega_\mathrm{rep}/2;\omega_\mathrm{rep}/2]$, corresponding to the shift of CEO induced by the difference phase velocity $\Omega_{\varphi, \sigma} = n \omega_\mathrm{ceo, \sigma}$.

Our initial goal is to demonstrate the two-dimensional frequency comb entirely through optical characterizations. We employ a \ce{Si3N4} microring resonator embedded in \ce{SiO2} with a core thickness of $H=770$~nm, a ring radius of $R=23$~{\textmu}m, and a ring width of $RW=1125$~nm. By injecting pump light with the transverse electric polarization into the microring at around 192~THz (1550~nm), we can produce a DKS with 150~mW of pump power on-chip and operating at $\omega_\mathrm{rep}\approx1016$~GHz, with thermal stabilization provied by a cooling laser at 306~THz. The secondary pump for the 2D frequency comb is injected 16 modes away from the primary pump at 208~THz (1442~nm) with 50~mW on-chip [\cref{fig:4}(a.ii)]. In good accord with FEM simulations that match the experimental $D_\mathrm{int}(\mu)$ measurement [\cref{fig:4}(a.i)], the secondary pumped component is shifted away from the DKS microcomb by about 3~GHz (about 25~pm at 1560~nm) as a result of the inherent resonator dispersion. The presence of additional characteristics that resemble dispersive waves distinguishes the obtained dual-pumped DKS frequency comb in the $\mu$ domain from the single-pumped one [\cref{fig:4}(a.ii)]. Each $\mu$ comb tooth may be decomposed into several sub-comb teeth owing to the cascading nonlinearity in the $\sigma$ domain. In addition to retrieving the envelope of each frequency comb component, the optical spectrum analyzer's resolution of 5~pm (630~GHz at 1550~nm), which is lower than $\Omega_\varphi\simeq D_\mathrm{int}(\mu_\mathrm{sp})\simeq4$~GHz, allows one to retrieve and recast the frequency comb in the $\{\mu,\Omega\}$ domain. Similar to the solitonic-OPO scenario, the obtained 2D-comb may be integrated along $\mu$, allowing us to resolve five distinct comb teeth along $\Omega$ and demonstrate an equal frequency spacing with $\Omega_\varphi = 4.1\;\mathrm{GHz}~\pm$. The multi-pumped LLE model produces a similar 2D comb [\cref{fig:4}(a.iv)] that, when the $\sigma$ comb teeth are extracted from the $\mu$ integration of the comb, highlights the excellent agreement between experiment and simulation [\cref{fig:4}(a.v)].

However, since not all $\mu$ modes will carry every $\sigma$ component owing to power concerns, it might be questioned whether the generated frequency comb represents a two-dimensional comb. One way to address this issue is to lower $\Delta\Omega$ as much as feasible. Since resonant enhancement happens for $D_\mathrm{int}\simeq0$, such a condition is satisfied at the DW. Using the microring resonator employed in the solitonic-OPO section, one may set the auxiliary pump precisely at the DW mode at $\mu=-92$ (194~THz) using the microring resonator employed in the solitonic-OPO section. Even while the created dual-pumped DKS frequency comb differs noticeably from the single-pumped one, especially with significantly higher power for $\mu<-50$, optical processing to observe the comb in the $\sigma$ domain is not achievable owing to the OSA resolution. We use a 6~GHz photodiode and a 6~GHz real-time electronic spectrum analyzer (RSA) to detect any CEO frequency offset over the whole frequency comb to resolve $\Delta\Omega$ and the various $\sigma$ comb teeth. We may infer that the many RSA-measured beat notes correspond to each $\sigma$ comb tooth from the 2D-comb and are separated by a set frequency $\Delta\Omega=1.132$~GHz~$\pm~2$~MHz (uncertainty coming from the standard deviation of the frequency separation of each teeth) since the all-optical measurement previously discussed emphasized the solitonic-OPO cascading process. The multi-pump single LLE allows us to explore the two-dimensional frequency comb shape, where not only the cascading solitonic-OPO occurs, creating multiple $\sigma$ comb teeth, but also underlines that each mode $\mu$ carries the other comb dimension. There are more comb teeth than just the handful that the limited photodetector and RSA bandwidth could measure in the $\Omega$ frequency domain. The LLE indicates that the system has 11 $\sigma$ teeth, and the $\mu$ comb spans over 200 modes, resulting in an actual two-dimensional comb.

\vspace{1ex}
\textbf{Four-wave mixing Bragg scattering in the phase velocity space -- }%
The two-dimensional frequency comb's cascaded form and the solitonic-OPO do not capture all possible wave mixings in a $\chi^{(3)}$ medium. Because of its similarity to the scattering of waves against a grating, one additional FWM process is given the name FWM-Bragg scattering (BS)~\cite{McKinstriePhys.Rev.A2012, LiNaturePhoton2016, SinghOptica2019,BellOptica2017}. In this FWM scenario, the effective grating is produced by two powerful pumps modulating the refractive index in the moving frame of the light through Kerr nonlinearity. As a result, the signal is scattered into a pair of idlers whose frequencies are shifted from the signal by the pump spacing. Thanks to its intrinsic noiselessness property and nearest neighbor coupling, this nonlinear mixing is crucial for applications like quantum frequency conversion~\cite{LiNaturePhoton2016} and the creation of synthetic frequency lattices~\cite{BellOptica2017, LiPhys.Rev.A2021, WangLightSciAppl2020}. Instead of coupling the various azimuthal modes $\mu$ in our system, we want to show that this specific and crucial nonlinear mixing process has an analog in the phase-velocity frequency space. We may first investigate this possibility using LLE simulation and the same microring as in the solitonic-OPO scenario (Fig.~\ref{fig:5}(a)). The DKS can scatter into other idlers thanks to two auxiliary pumps at $\mu=-87$ and $\mu=81$. The pump power controls whether the procedure can continue to cascade, creating a frequency comb in the $\Omega$ dimension with a frequency spacing driven by that between the two pumps. The nonlinear coupling, however, is distinct from the reported 2D comb above. In fact, \cref{eq:2} demonstrates that an all-to-all coupling exists, in contrast to the cascaded FWM-BS, which exclusively couples the closest neighbors to produce a synthetic frequency lattice. By adding multiple auxiliary pumps to the frequency comb devices we have thus far studied, we may experimentally study this FWM-BS process. The $\Omega$ frequency space should first be probed when only one auxiliary pump is present. As demonstrated for three different auxiliary pumped modes, this wavepacket color is frequency shifted from the DKS by $\Delta\Omega\simeq D_\mathrm{int}(\mu_\mathrm{sp})$, as expected [\cref{fig:5}(b)]. When introducing two auxiliary pumps, following each combination presented before, one could notice that the frequency comb in the $\{\mu, \Omega\}$ space presents new idlers [\cref{fig:5}(c)]. One may observe that, after integrating along subsets of $\mu$ for filtering purposes to obtain $S_\Omega$, the produced new idler is at a $\Omega$ frequency equal to the difference between the two auxiliary pumps, as expected from the FWM-BS process. LLE simulation of the experiment is in excellent agreement with the observation, which is limited by the resolution of the OSA. It is interesting to observe that FWM-BS produces many idlers for the auxiliary pump $\mu=-88$ and $\mu=-57$ combination. The DKS from this nonlinear mixing gives rise to the first idler directly. The second one, in contrast, stresses the potential for cascading this process by matching to another FWM-BS between the secondary pump and the first idler.

\vspace{1ex}
\textbf{Discussion -- } %
A DKS phase velocity and group velocity mismatch results in the well-known carrier envelope offset, ubiquitous in all frequency comb work. However, until now, this degree of freedom was only considered in terms of its consequence for locking the DKS while discarding potential nonlinear mixing. Because the CEO frequency is, per definition, bounded between $[-\omega_\mathrm{rep}/2; +\omega_\mathrm{rep}/2]$, any new light nonlinearly produced would be in this phase-velocity dimension frequency and shifted by orders of magnitude lower than the repetition rate. We have shown that when an auxiliary pump is used, the cavity's intrinsic dispersion against the DKS dispersion-less property enables harnessing of a phase-velocity shift that ignites the nonlinearity in this dimension. We demonstrate that this phenomenon may be described using a set of coupled LLEs allowing for phase-matching criteria, much like mode mixing. Through experiments, we show that any FWM process may be used in the phase-velocity dimension, which is, by definition, orthogonal to the frequency comb's mode number dimension. As a result, we have demonstrated both theoretically and experimentally that a two-dimensional frequency comb may form, producing two repetition rates that are three orders of magnitude apart. With this proof of concept, the microwave and optical domains may be further connected without the use of a low repetition rate resonator, enabling new opportunities for connecting the optical and microwave domains using DKS microcombs.


\bibliographystyle{naturemag}

\section*{Methods}
\textbf{LLE simulations -- }
For the closest simulation of the experimental data, we aim to reproduce numerically exactly what the experiment is: we normalize the LLE with $\omega_\mathrm{rep}$ from the DKS, extract at each round-trip the envelope recreating a pulse train which we can Fourier transform. From this obtianed frequency comb, we then processed it exactly as the experimental one, recreating the $\Omega$ second dimension from the repetition rate and each $\mu$ component enabling us for a direct comparison with experiment. 

\vspace{1ex}
\noindent \textbf{\large Data availability} \\
The data that supports the plots within this paper and other findings of this study are available from the corresponding authors upon reasonable request.

\vspace{1ex}
\noindent \textbf{\large Author Contributions} \\
G.M. led the project, designed the ring resonators, conducted the experiments, helped developed the theoretical framework, developed the simulation tools and perform the modeling. C.M. contributed in the understanding of the physical phenomenon, with the help of P.S, by developing the theoretical model. K.S. helped with the data processing and understanding along with A.D and Y.K.C. C.L, J.S and, M.C  helped with the characterizations. G.M. and K.S. wrote the manuscript, with input from all authors. All the authors contributed and discussed the content of this manuscript.

\vspace{1ex}
\noindent \textbf{\large Acknowledgements} \\
The photonics chips where fabricated by Ligentec SA. The authors acknowledge partial funding support from the AFRL Space Vehicle Directorate, the DARPA APHI program, and the NIST-on-a-chip program.

\vspace{1ex}
\noindent \textbf{\large Competing Interests} \\
The authors declare no competing interests.



\end{document}